\documentclass[
]{ceurart}

\usepackage{listings}
\begin{document}

\copyrightyear{2023}
\copyrightclause{Copyright for this paper by its authors.
  Use permitted under Creative Commons License Attribution 4.0
  International (CC BY 4.0).}

\conference{The 4th Workshop on Scientific Document Understanding,  February 26, 2024, Vancouver, BC, Canada}

\title{Sparse Meets Dense: A Hybrid Approach to Enhance Scientific Document Retrieval}


\author[]{Priyanka Mandikal}[%
email=mandikal@utexas.edu,
url=https://priyankamandikal.github.io/,
]
\address[]{The University of Texas at Austin, USA}

\author[]{Raymond Mooney}[%
email=mooney@utexas.edu,
url=https://www.cs.utexas.edu/~mooney/,
]


\begin{abstract}
Traditional information retrieval is based on sparse bag-of-words vector representations of documents and queries. More recent deep-learning approaches have used dense embeddings learned using a transformer-based large language model. We show that on a classic benchmark on scientific document retrieval in the medical domain of cystic fibrosis, that both of these models perform roughly equivalently. Notably, dense vectors from the state-of-the-art SPECTER2 model do not significantly enhance performance. However, a hybrid model that we propose combining these methods yields significantly better results, underscoring the merits of integrating classical and contemporary deep learning techniques in information retrieval in the domain of specialized scientific documents.
\end{abstract}


\maketitle

\section{Introduction}

The field of information retrieval (IR) has undergone significant transformations over the years, evolving from traditional methods based on sparse vector representations to more advanced techniques utilizing deep learning models. Historically, the vector-space model in IR, as described by Manning et al.~\cite{manning:book08}, has been predominant. This model represents documents and queries using sparse, Term-Frequency/Inverse-Document-Frequency (TF/IDF) weighted bag-of-words (BOW) vectors, a method known for its simplicity and effectiveness in various text retrieval applications.

With the advent of deep learning, there has been a paradigm shift towards utilizing dense embeddings, especially those derived from transformer-based large language models (LLMs). These methods, exemplified by Dense Passage Retrieval (DPR)~\cite{dpr}, offer a more nuanced and context-aware approach to text representation, often resulting in improved retrieval performance compared to traditional sparse-vector methods.

In the specialized field of scientific document retrieval, dense-vector approaches have gained prominence. One notable example is the SPECTER model~\cite{specter}, which uses a BERT-based framework for embedding scientific texts. It employs contrastive learning, leveraging citation information for supervision, to identify document similarities effectively. 
On a variety of tasks including document classification, citation prediction, and recommendation, SPECTER has been shown to outperform a variety of competing methods~\cite{specter}.
The model's recent iteration, SPECTER2~\cite{singh2023scirepeval}, further refines this approach by training on an expanded corpus of 6M triplets spanning a wider array of scientific fields, producing more representative document embeddings.

In this paper, 
we evaluate SPECTER2 on a classic benchmark in scientific document retrieval in the medical area of cystic fibrosis~\cite{cf-data}.
Contrary to the anticipated superiority of dense embeddings, our findings reveal that the performance of SPECTER2 is comparable, and in some cases, marginally inferior to that of traditional sparse-vector retrieval methods. This observation led us to explore a hybrid retrieval model that combines the strengths of both sparse and dense vector representations. 
This simple, yet elegant hybrid approach performs remarkably well, clearly outperforming the base models on both standard precision/recall and NDCG metrics.   
Through our research, we aim to determine the optimal blend of these approaches and demonstrate the potential benefits of such a hybrid model in enhancing document retrieval accuracy and efficiency.

Our work contributes to the evolving landscape of IR by not only comparing traditional and modern IR techniques but also by demonstrating the efficacy of a hybrid approach in a specialized medical domain. 
This study not only contributes to the existing body of knowledge in IR but also provides practical insights for the development of more effective retrieval systems, particularly in specialized fields such as medical research. The findings have implications for both the theoretical understanding of IR models and their application in real-world scenarios.

\section{Related Work}

\paragraph{Classic IR.} The evolution of information retrieval (IR) techniques, particularly in the context of scientific document retrieval, has seen a significant shift from traditional sparse vector models to sophisticated deep learning approaches. The seminal work of Salton and McGill~\cite{salton1983introduction} laid the foundation for vector-space models, employing bag-of-words (BOW) representations and TF/IDF weighting, a standard in early IR systems. This approach has been extensively studied and optimized over the years, as documented by Manning et al.~\cite{manning:book08}.

\vspace{-0.25em}
\paragraph{Deep learning based IR.} The advent of deep learning has introduced a new paradigm in IR, with dense vector embeddings offering nuanced text representations. Pioneering this shift, Le and Mikolov~\cite{le2014distributed} demonstrated the effectiveness of dense embeddings with their work on Doc2Vec, showing significant improvements in capturing semantic text similarities. 

\vspace{-0.25em}
\paragraph{Large Language Models.} The development of transformer-based models, as introduced by Vaswani et al.~\cite{vaswani2017attention}, further revolutionized this field, enabling more complex and contextually rich text representations. Focusing on scientific documentation, the SPECTER model~\cite{specter}, based on BERT architecture~\cite{devlin2019bert}, emerged as a significant advancement, utilizing citation networks for improved document embeddings. Its successor, SPECTER2~\cite{singh2023scirepeval}, expanded on this with a larger training corpus and enhanced learning strategies.
Recent research has also explored the integration of sparse and dense vector methods~\cite{arivazhagan2023hybrid,luan2021tacl}. However, such a hybrid approach has not been previously explored for scientific document retrieval. Such hybrid models are particularly relevant in scientific domains where traditional IR techniques remain competitive, as observed in our study focusing on cystic fibrosis document retrieval.

\section{Dataset}

The Cystic Fibrosis Database (CF)~\cite{cf-data} consists of 1,239 documents published from 1974 to 1979 discussing Cystic Fibrosis, and a set of 100 queries with the respective relevant documents as answers. It is a classic IR benchmark that is discussed in an early textbook~\cite{baeza-yates:book99}. The documents are represented by an abstract and lists of minor and major subjects, but do not include the full text of the document. The queries are full sentence questions, e.g. ``Can one distinguish between the effects of mucus hypersecretion and infection on the submucosal glands of the respiratory tract in CF?'' Each query has been annotated with ratings from four domain experts, where each document is assigned a relevance score: 0: not relevant, 1: marginally relevant, 2: highly relevant.  

When measuring precision and recall, any document rated as at least marginally relevant by at least one annotator is considered relevant. To provide a more nuanced gold-standard, the average of the judges' ratings is also used as a continuous relevance value for each document. This is used when measuring Normalized Discounted Cumulative Gain (NDCG), a standard IR metric that utilizes continuous gold-standard relevance ratings~\cite{ndcg,manning:book08}.

\section{Approach}

\begin{figure*}[!t]
\centering
\begin{center}
\includegraphics[width=\linewidth]{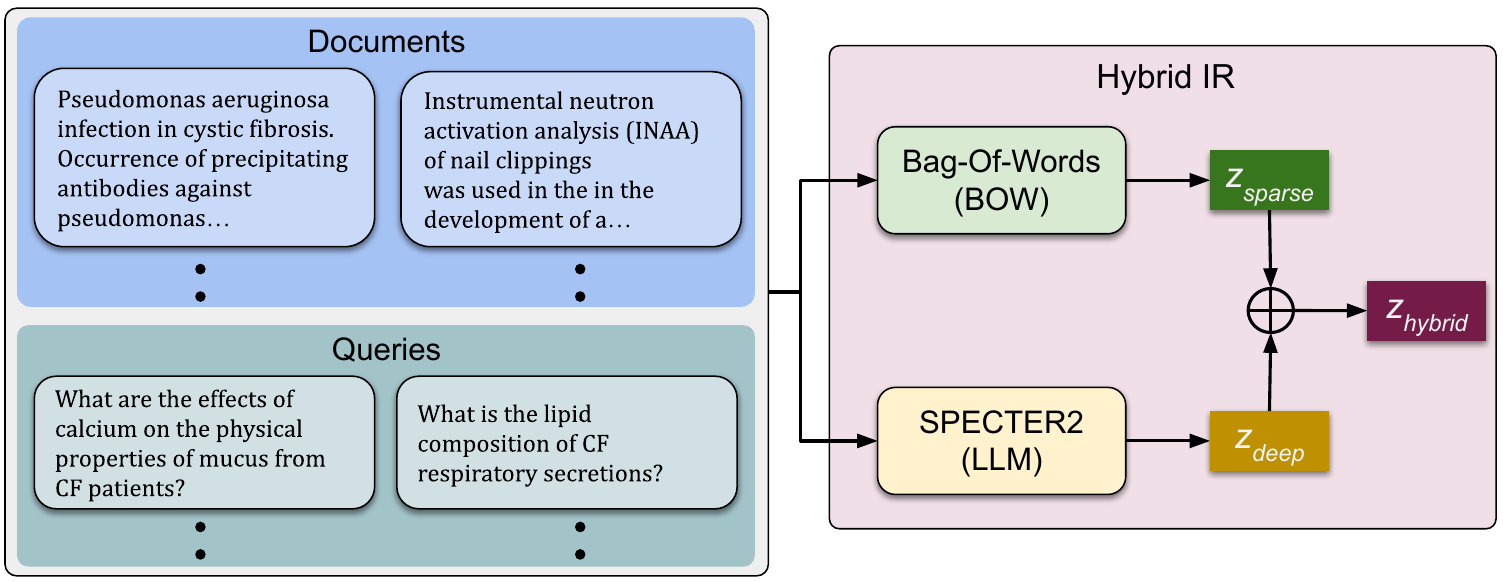}
\end{center}
\caption{\textbf{Overview of our approach.} On a medical dataset of cystic fibrosis documents, we combine sparse bag-of-words embeddings with dense embeddings from a SOTA LLM (Specter2~\cite{singh2023scirepeval}) to produce a hybrid retriever that significantly outperforms both methods. }
\label{fig:overview}
\end{figure*}

We present an overview of our hybrid retriever combining sparse and deep embeddings in Fig.~\ref{fig:overview} and describe each component below.

\subsection{Sparse Retrieval Model}

To generate sparse embeddings, we use a simple, classical, bag-of-words vector-space retrieval (VSR)  method as a baseline.  It uses Term-Frequency/Inverse-Document-Frequency (TF/IDF) token weighting and cosine similarity to measure query/document relevance~\cite{baeza-yates:book99,manning:book08}.  A traditional inverted index is used to efficiently retrieve documents that share tokens with the query after removing a standard set of stopwords.  

\subsection{Dense Retrieval Model}

To generate deep embeddings, we use SPECTER2~\cite{singh2023scirepeval} as a representative of a state-of-the-art deep-learning model explicitly trained to produce effective dense-vector representations of scientific documents.  SPECTER2 is a successor to SPECTER~\cite{specter} which in turn was a followup to SciBERT~\cite{scibert}.  It uses a transformer-based LLM that was explicitly trained on a large corpus of scientific documents from the Semantic Scholar search engine~\cite{semanticscholar}.

SPECTER2 uses additional self-supervised training based on citations to produce good document abstract embeddings. A contrastive learning approach is used to encourage similar 768-dimensional embeddings for an article its direct citations. SPECTER2 has been trained on 6M triplets spanning 23 fields of study. Our retrieval model ranks documents based on a similarity score between the dense vectors generated for a query and the document abstract. 

\begin{figure*}
  \centering
  \begin{minipage}{.48\linewidth}
    \centering
    \includegraphics[clip, trim=1.8cm 1.5cm 9cm 6cm, width=\linewidth]{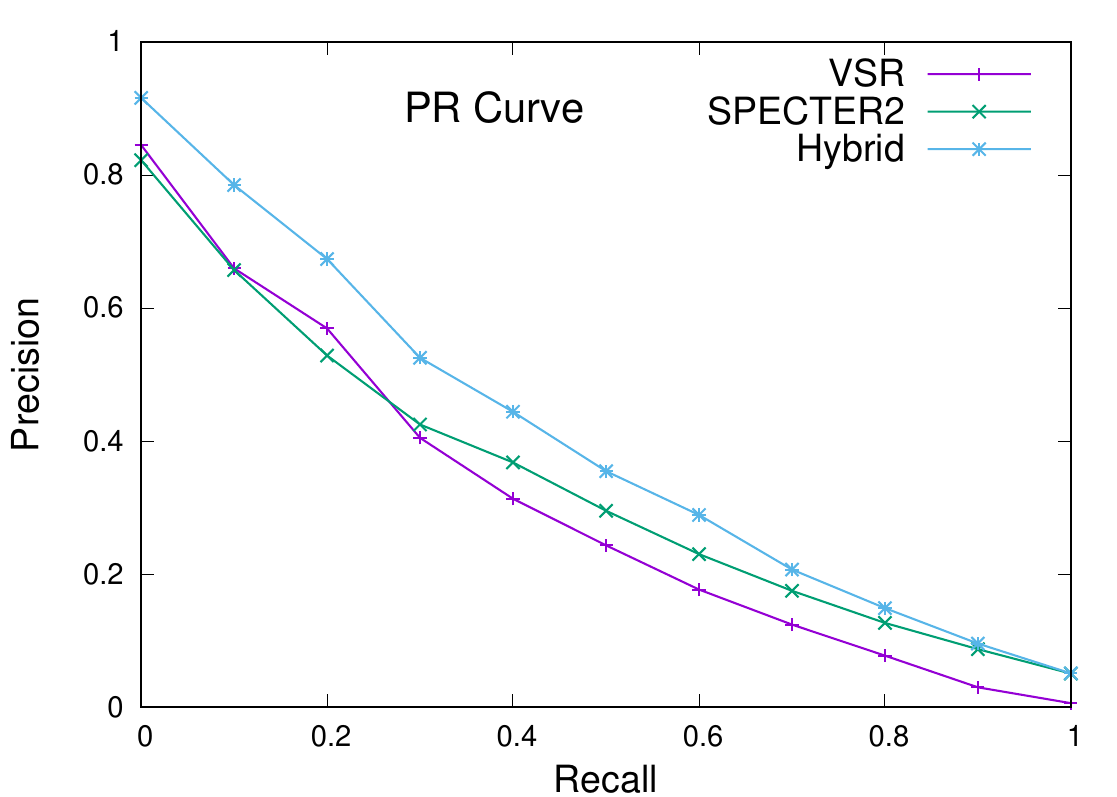}
  \end{minipage}%
  \hspace{0.4cm}
  \begin{minipage}{.48\linewidth}
    \centering
    \includegraphics[clip, trim=1.8cm 1.5cm 9cm 6cm, width=\linewidth]{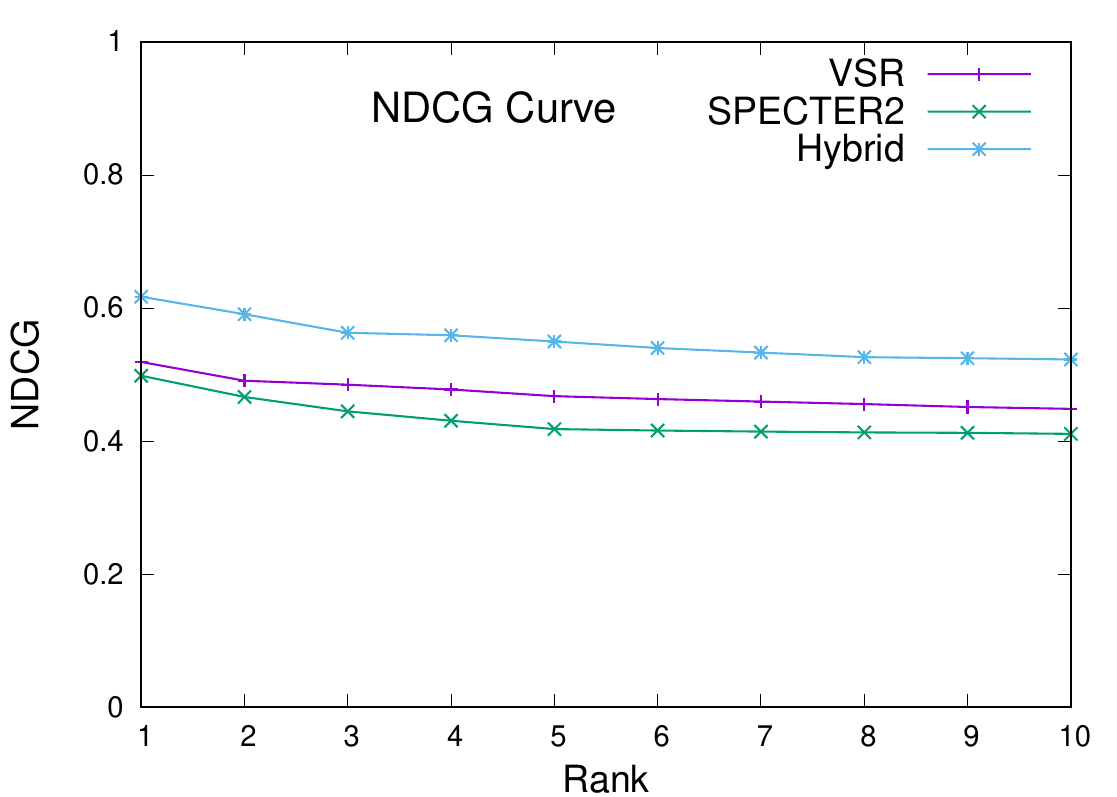}
  \end{minipage}
  \caption{\textbf{Results on the Cystic-Fibrosis dataset.} The hybrid approach ($\lambda=0.8$) outperforms both traditional sparse vector-space retrieval (VSR) and state-of-the-art deep embeddings (SPECTER2~\cite{singh2023scirepeval}) in both PR (left) as well as NDCG (right) metrics.}
  \label{fig:results_main}
\end{figure*}

\begin{figure*}[!h]
\centering
\includegraphics[width=\linewidth]{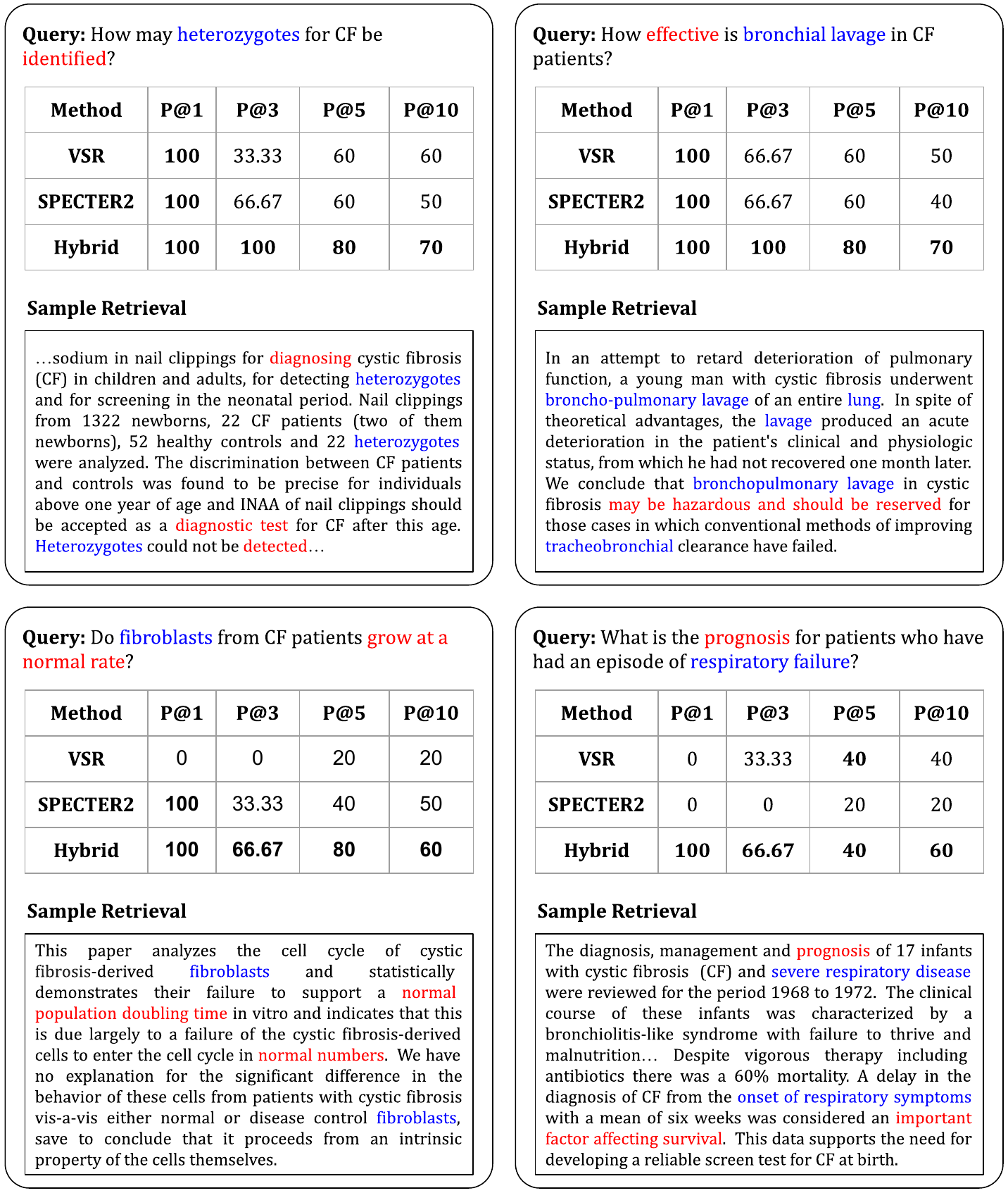}
\caption{\textbf{Sample queries and retrievals.} We show results for the three methods on four sample queries. Our hybrid approach outperforms both VSR and SPECTER2, retrieving more number of relevant documents higher in the retrieval ranks. We also show sample retrievals relevant to the given query with keywords highlighted.}
\label{fig:sample_retrievals}
\end{figure*}

\subsection{Hybrid Retrieval Model}

Finally, we combine the sparse and dense retrieval models to produce a hybrid retriever (see Fig.~\ref{fig:overview}). Our hybrid model uses a simple weighted combination of the query/document similarities in the sparse and dense embedding spaces. Specifically, we score a document $D$ for query $Q$ using the ranking function:
\begin{equation}
  \begin{aligned}  
  S_{hybrid}(D,Q) = & \lambda \ Sim\big(z_{dense}(D), z_{dense}(Q)\big) \ + \\ 
  & (1 - \lambda) \ Sim\big(z_{sparse}(D), z_{sparse}(Q)\big)
  \end{aligned}
\end{equation}
Where $z_{dense}$ and $z_{sparse}$ denotes the dense and sparse embedding functions and $Sim$ is cosine similarity measuring the angle between such vector embedddings.
In our experiments we present results for different values of the hyperparameter $\lambda$.

\section{Experiments}

In this section, we provide empirical results for the effectiveness of hybrid embedding and also perform ablations on various hyperparameter and model variants. 

\paragraph{Effectiveness of hybrid embeddings.}
We compare the sparse, dense and hybrid retrieval models on the CF corpus using standard IR experimental methods \cite{manning:book08}.
We plot precision-recall (PR) curves using a standard approach that interpolates retrieval precision for a set of standard recall levels \cite{baeza-yates:book99}.  We also compute NDCG values \cite{ndcg} for the top 10 retrievals using the continuous gold-standard relevance ratings determined from the average human evaluator ratings.
We use cosine similarity as our similarity metrics while comparing the document and query embeddings.
We also tried using Euclidean distance to compare deep embeddings, but it performed almost identically to cosine similarity.

Quantitative results are shown in Fig.~\ref{fig:results_main}. We observe that the hybrid approach ($\lambda=0.8$) outperforms both traditional sparse embeddings (VSR) and state-of-the-art deep embeddings (SPECTER2~\cite{singh2023scirepeval}) in both PR as well as NDCG metrics. 
We also show results for four sample queries in Fig.~\ref{fig:sample_retrievals}. We can see that while both VSR and SPECTER2 fail to consistently retain high precision values for their retrievals, our hybrid approach is able to effectively retrieve greater number of highly relevant documents higher up in the retrieval list. We also depict sample retrievals with keywords highlighted for the given queries to illustrate the relevance of the retrieved documents to the user query.
This underscores the simple yet effective strategy of combining sparse and dense embeddings for scientific document retrieval.

\begin{figure*}
  \centering
  \begin{minipage}{.48\linewidth}
    \centering
    \includegraphics[clip, trim=1.8cm 1.5cm 9cm 6cm, width=\linewidth]{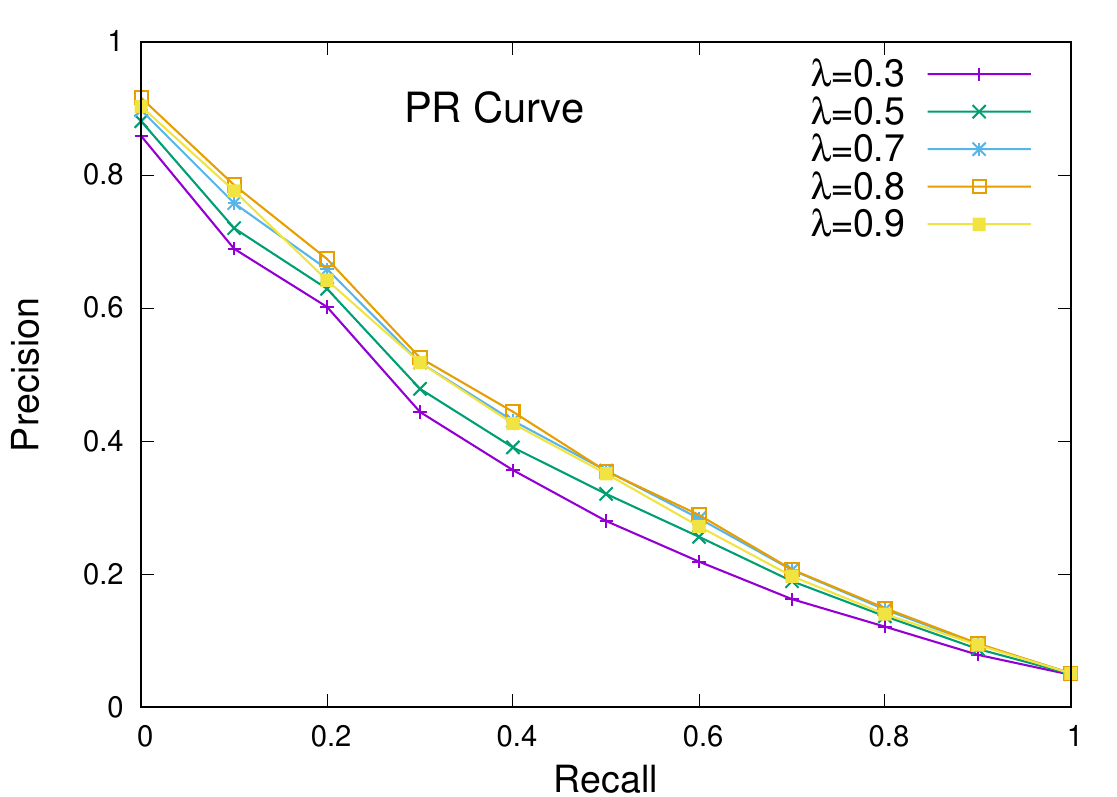}
  \end{minipage}%
  \hspace{0.4cm}
  \begin{minipage}{.48\linewidth}
    \centering
    \includegraphics[clip, trim=1.8cm 1.5cm 9cm 6cm, width=\linewidth]{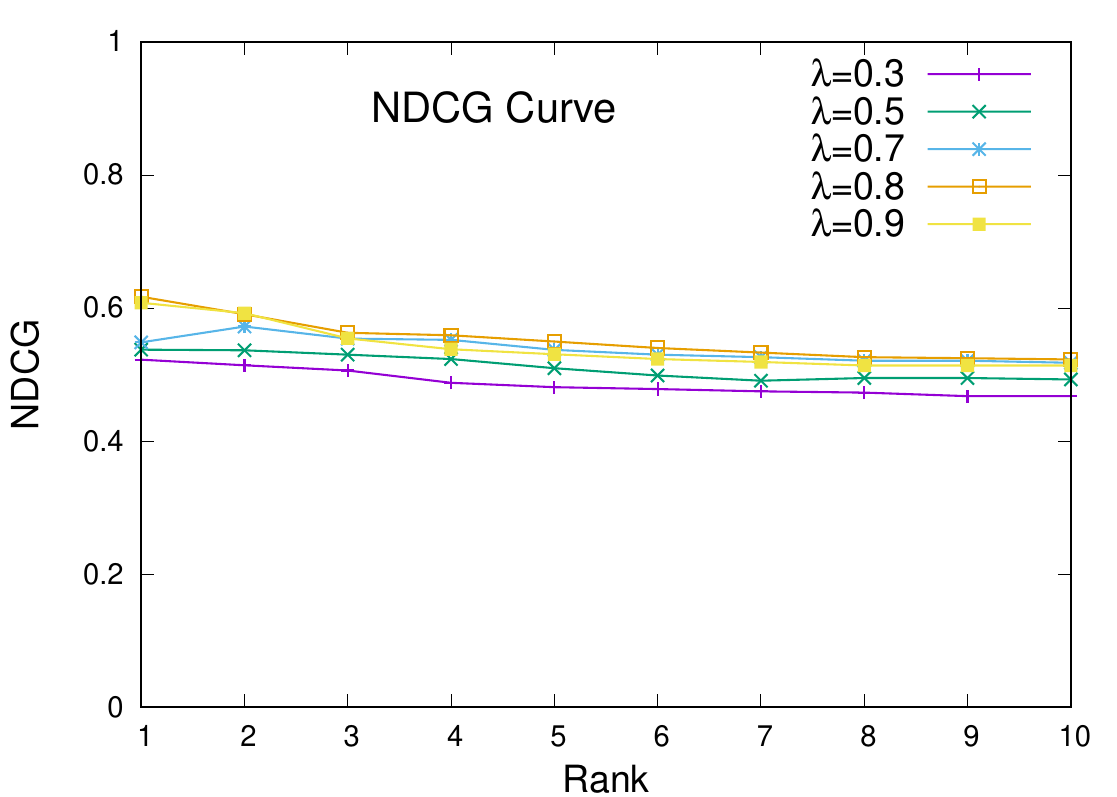}
  \end{minipage}
  \caption{\textbf{Ablations with different values of $\lambda$.} Weighing the deep embedding with a weight of $\lambda=0.8$ (where sparse gets a weight of $0.2$) produces the best results for both PR and NDCG metrics.}
  \label{fig:results_lambda}
\end{figure*}

\begin{figure*}
  \centering
  \begin{minipage}{.48\linewidth}
    \centering
    \includegraphics[clip, trim=1.8cm 1.5cm 9cm 6cm, width=\linewidth]{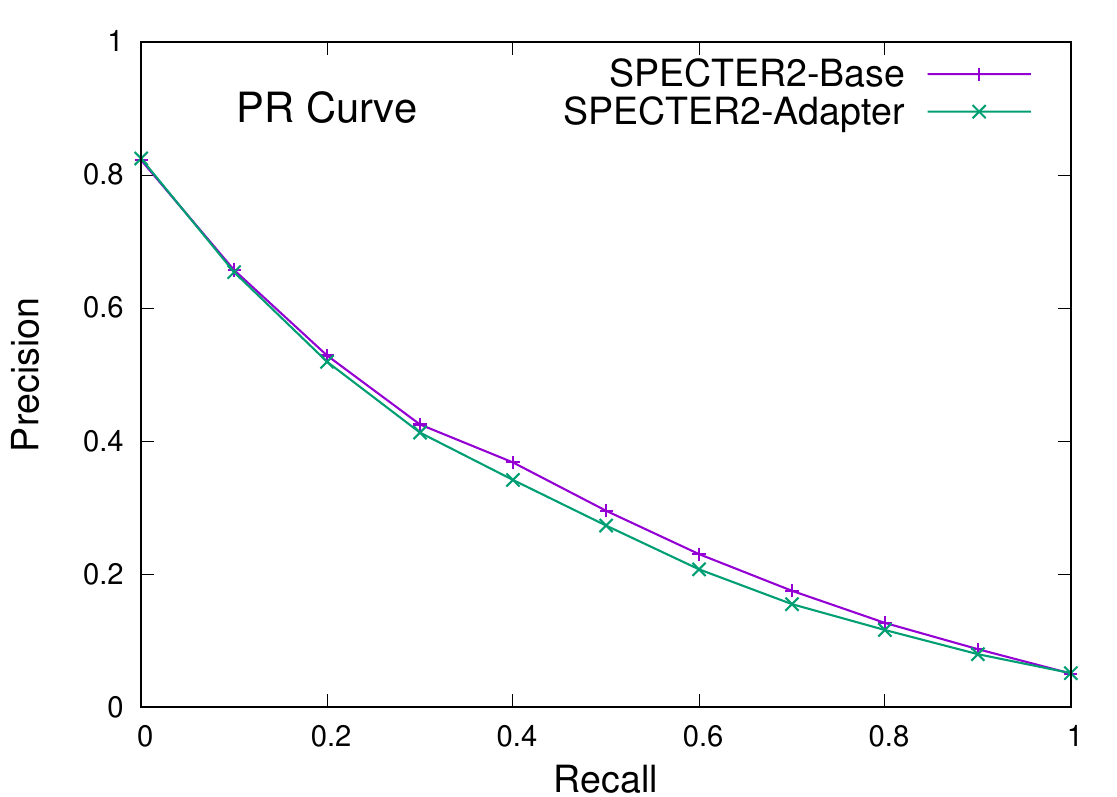}
  \end{minipage}%
  \hspace{0.4cm}
  \begin{minipage}{.48\linewidth}
    \centering
    \includegraphics[clip, trim=1.8cm 1.5cm 9cm 6cm, width=\linewidth]{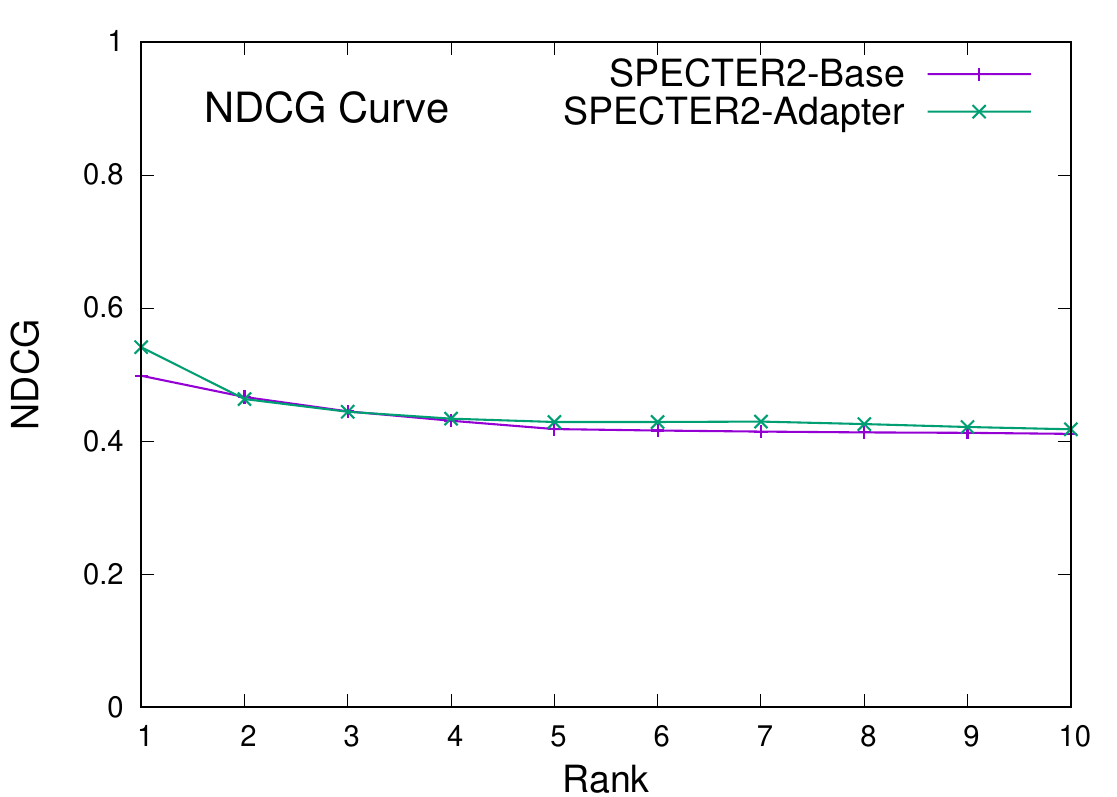}
  \end{minipage}
  \caption{\textbf{Ablations with SPECTER2 base vs adapter versions.} The base model performs as well as the adapter variant on the PR metric. The adapters marginally improve performance on NDCG but hurt precision for high recall levels .}
  \label{fig:results_specter}
\end{figure*}

\paragraph{Hyperparameter $\lambda$.}
We conduct ablations on the value of $\lambda$ for effective combination of the dense and sparse embeddings. 
Results are depicted in Fig.~\ref{fig:results_lambda}. We observe that weighing the deep embedding with a weight of $\lambda=0.8$ (where sparse gets a weight of $0.2$) produces the best results for both PR and NDCG metrics. 
This indicates that a higher emphasis on the dense embeddings from the SPECTER2 model, while still incorporating a significant contribution from the traditional sparse BOW vectors, achieves an optimal balance for document retrieval in the cystic fibrosis dataset. This hybrid approach effectively captures the strengths of both methodologies, utilizing the contextual richness and semantic understanding from the dense embeddings, and the keyword-focused precision of sparse vectors. Therefore, our findings underscore the importance of a nuanced combination of these models, tailored to the specific characteristics of the dataset and the retrieval task at hand. While scaling this hybrid approach to different datasets, this hyperparameter can be set using normal grid-search on a validation set.

\paragraph{Base vs Adapters.}
SPECTER2 includes adapters that are explicitly trained to produce improved representations of abstracts and queries, but we found that they did not improve performance over the base model on the cystic-fibrosis dataset. This ablation is shown in Fig.~\ref{fig:results_specter}. The base model performs as well as the adapter variant on the PR metric. The adapters only marginally improve performance on NDCG while marginally decreasing  precision at higher recall levels. Therefore, our results in Figs.~\ref{fig:results_main} and ~\ref{fig:results_lambda} utilize the SPECTER2 base model.

\section{Conclusion}

This paper has explored applying SOTA deep embeddings for scientific documents using the SPECTER2 model to a classic benchmark dataset in the domain of medical document retrieval.  We found that they gave disappointing results, at best performing only about as well as traditional sparse TF/IDF vector-space retrieval. However, a simple hybrid approach that combines sparse and dense retrieval gave significantly improved results.  
The findings align with the growing consensus in the literature that the future of IR lies in the thoughtful integration of classical and modern methodologies.


\bibliography{sample-ceur}

\appendix

\end{document}